\documentclass[a4paper]{article}

\usepackage{multirow}
\usepackage{INTERSPEECH2021}
\usepackage[marginal]{footmisc}
\renewcommand{\footnote}{}

\title{The SpeakIn System for VoxCeleb Speaker Recognition Challange 2021}
\name{Miao Zhao\textsuperscript{*}, Yufeng Ma\textsuperscript{*}, Min Liu, Minqiang Xu\textsuperscript{\dag}}
\address{
  SpeakIn Technologies Co. Ltd.}
\email{\{zhaomiao, mayufeng, liumin, xuminqiang\}@speakin.ai}

\begin{document}

\maketitle
\begingroup\renewcommand\thefootnote{*}
\footnotetext{These authors share equal contribution to this work.}
\begingroup\renewcommand\thefootnote{\dag}
\footnotetext{Corresponding author.}
\begingroup\renewcommand\thefootnote{1}
\footnotetext{https://github.com/kaldi-asr/kaldi/tree/master/egs/voxceleb/v2}

\begin{abstract}
  This report describes our submission to the track 1 and track 2 of the VoxCeleb Speaker Recognition Challenge 2021 (VoxSRC 2021). Both track 1 and track 2 share the same speaker verification system, which only uses VoxCeleb2-dev as our training set. This report explores several parts, including data augmentation, network structures, domain-based large margin fine-tuning, and back-end refinement. Our system is a fusion of 9 models and achieves first place in these two tracks of VoxSRC 2021. The minDCF of our submission is 0.1034, and the corresponding EER is 1.8460\%.
\end{abstract}

\section{System Description}

For both Track 1 and Track 2, we adopt the same system settings without any extra data other than Voxceleb2-dev \cite{chung2018voxceleb2}. This part will focus on the method we implemented in this challenge.

\subsection{Datasets and Data Augmentation}
\subsubsection{Training Data}\label{data_sec}
The VoxCeleb2-dev dataset contains 1,092,009 utterances and 5,994 speakers in total. Data augmentation is also quite important in training speaker verification models.  We here adopted a 3-fold speed augmentation \cite{yamamoto2019speaker, wang2020dku} at first to generate extra twice speakers. Each speech segment in this dataset was perturbed by 0.9 or 1.1 factor based on the SoX speed function. Then we obtained 3,276,027 utterances and 17,982 speakers. The traditional Kaldi-based \cite{povey2011kaldi, snyder2018x} method (offline augmentation) is widely adopted in this field. Recent researches \cite{cai2020fly, ravanelli2021speechbrain} mentioned a new method that augments data on the fly (online augmentation). Our system contains both offline and online trained models. These two different data augmentation methods are applied separately for different training modes:

\begin{itemize}
 \item \textbf{Offline training mode}: In this training method, we used RIRs \cite{ko2017study} and MUSAN \cite{snyder2015musan} to create extra four copies of the training utterances and the data augmentation process was based on the Kaldi VoxCeleb recipe\textsuperscript{1}. After this augmentation, 16,380,135 utterances from 17,982 speakers were generated to extract acoustic features.
 \item \textbf{Online training mode}: Instead of concatenating different types of augmentation \cite{ravanelli2021speechbrain}, we adopted a chain-like augment. It means that we predefine an effect chain composed of several augments, and every augment should have its probability to be activated. The effect chain is as:
\begin{itemize}
    \item gain augment with a probability of 0.2
    \item white noise augment with a probability of 0.2
    \item RIR reverberation and noise addition augment with a probability of 0.6
    \item time stretch augment with a probability of 0.2
\end{itemize}
It is worth mentioning that the offline 3-fold speed augmentation is also adopted in online augmentation, which means the number of classes is 17,982. The speed augmentation will change the pitch of a speaker, while time-stretching will not change the pitch. Both foreground and background noises are added, and they are randomly selected from MUSAN and RIRs noises.
\end{itemize}

\subsubsection{Developing Set}
To evaluate the performance of our models, we used 5 test sets \cite{chung2018voxceleb2,nagrani2017voxceleb} as our developing sets: 
\begin{itemize}
 \item \textbf{VoxCeleb1-O}: 37,720 trails are sampled from the VoxCeleb1 test dataset with only 40 speakers.
 \item \textbf{VoxCeleb1-E}: This is an extended version of VoxCeleb1-O. This set contains 581,480 trials from 1251 speakers.
 \item \textbf{VoxCeleb1-H}: This set has 552,536 trials. It is harder since each pair in this set shares the same nationality and gender.
 \item \textbf{VoxSRC20-dev}: This is the validation set of VoxSRC2020 and the trials contains out-of-domain
 data provided by VoxCeleb\_cd. This set has 263,486 trials.
 \item \textbf{VoxSRC21-val}: This is the validation set of VoxSRC2021 and has 60,000 trials. Trials in this set contain more multi-lingual data.
\end{itemize}

\subsubsection{Features}
We extracted both 81-dimensional and 96-dimensional log Mel filter bank energies based on Kaldi in offline training mode. The window size is 25 ms, and the frameshift is 10 ms. 200 frames of features were extracted without extra voice activation detection (VAD).  The speech segments were sliced to 2 seconds and augmented on the fly in the online training mode. 96-dimensional log Mel filter bank energies were extracted based on torchaudio. All features were cepstral mean normalized in both our training modes.

\subsection{Network Structures}


\subsubsection{Backbone}
\begin{figure}[t]
  \centering
  \includegraphics[width=0.48\textwidth,height=0.2\textwidth]{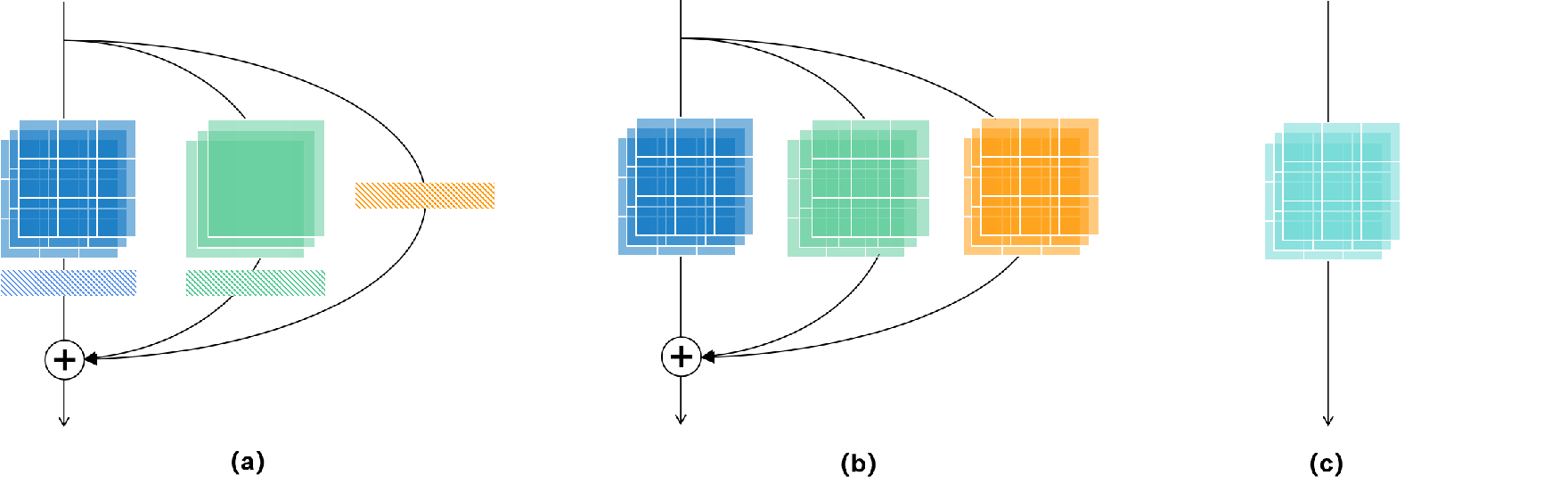}
  \caption{\textbf{Architecture of RepVGG block}. Here (a) is the training time state. (b) demonstrates the process of conv-bn fusion. (c) is the inference time state. $\oplus$ denotes element-wise addition.}
  \label{repvgg_block}
\end{figure}

Convolutional Neural Networks \cite{cai2020fly, thienpondt2021idlab, zeinali2019but} have become the main-stream solution in speaker verification tasks. Our backbones include two types of state-of-the-art models:

\begin{itemize}
    \item \textbf{RepVGG} \cite{ding2021repvgg} Recent researches proposed a new way to construct ConvNets. The method is called the re-parameterization technique. This method decouples the training time and inference time architecture. RepVGG, as one of the re-parameterized models, shows competitive performance in the computer vision field. We, at the first time, introduced this RepVGG architecture in speaker verification. As Figure \ref{repvgg_block} shows, the RepVGG block has a separate 3x3 and 1x1 convolutional layer with batch normalization and an identity branch with only a batch normalization layer during the training time. Since convolution and batch normalization can fuse into a convolution layer and both the 1x1 convolution layer and the batch normalization layer can transform to a 3x3 convolution layer, all branches in this block are equal to three 3x3 convolutions. All these 3x3 convolutions share the same setting (kernel size, stride, groups, dilation, and so on) so that they can fuse into only one 3x3 convolution by simply adding parameters filter-wisely. When merged into one 3x3 convolution and a ReLU layer, this block is as same as a VGG block during the inference time. We select RepVGG-A2, RepVGG-B1, RepVGG-B2g4, and RepVGG-B2 as our backbones. All models adopt 64 base channels except RepVGG-A2 which uses 96 base channels.
    \item \textbf{ResNet} \cite{he2016deep} As one of the most classical ConvNets, ResNet has proved its power in speaker verification. In our systems, both basic-block-based ResNet-34 and bottleneck-block-based ResNet (deeper structures: ResNet-101 and ResNet-152) are adopted. All base channels of these ResNets are 64.
\end{itemize}

\subsubsection{Pooling Method}
The pooling layer aims to aggregate the variable sequence to an utterance level embedding. The vanilla idea to achieve this purpose is by calculating the mean and standard derivation along the time axis \cite{snyder2017deep}. However, it could be limited by the fact that the contributions from different frames could be unequal. An attention mechanism  \cite{okabe2018attentive} is introduced to calculate weighted statistics of the outputs of the backbone. Furthermore, a multi-head mechanism was introduced to increase the diversity of attention, such as multi-head self-attentive (MHSA) pooling \cite{zhu2018self} and self multi-head attention (MHA) pooling \cite{india2019self}. The main difference between these two methods is the definition of the heads in attention mechanism. Instead of attending to the whole feature through different heads as we called queries, the latter split the features into several parts, and each head focuses on its corresponding part. We proposed a multi-query multi-head attention pooling mechanism (MQMHA) for the first time by combining both the multi-head methods above. Since this method can help us attend to different parts and gain more diversified information. The method can be described as below:

Suppose we have a backbone output $O=[o_1,o_2,...,o_T]$, $o_t\in \mathbb{R}^d$ and each $o_t$ is spit into $H$ parts with $o_t=[o_{t}^{1},o_{t}^{2},...,o_{t}^{H}]$, where $H$ is the number of head of attention. For each head, it has $Q$ trainable query vectors where $\mu_{h}^{q}\in\mathbb{R}^{d/H}$ . Attention weight of $w_{t,h,q}$ is defined as:

\begin{equation}
  w_{t,h,q} = \frac{\mathrm{exp}((o_{t}^{h})^T\mu_{h}^{q})}{\sum^T_{j=1}\mathrm{exp}\,((o_{j}^{h})^T\mu_{h}^{q})}
  \label{eq1}
\end{equation}
And the representation is expressed as:

\begin{equation}
  m_{h,q} = \sum_{i=1}^{T}(o_{i}^{h})^T w_{t,h,q}
  \label{eq2}
\end{equation}
as the MQMHA combines both MHSA and MHA, in which $H=1, Q>1$ and $H>1, Q=1$ are the cases of MHSA and MHA respectively.

Finally, we concatenate all of the sub-representations to get the utterance level embedding with $E_m=[\hat{m}_{1},\hat{m}_{2},...,\hat{m}_{H}]$, where $\hat{m}_{h}=[m_{h}^{1},m_{h}^{2},...m_{h}^{Q}]$. And an extra attentive standard deviation $E_{std}$ computed through the attention weights. This standard deviation is concatenated with $E_m$ to enhance the performance.

\begin{figure}[t]
  \centering
  \includegraphics[width=0.45\textwidth,height=0.2\textwidth]{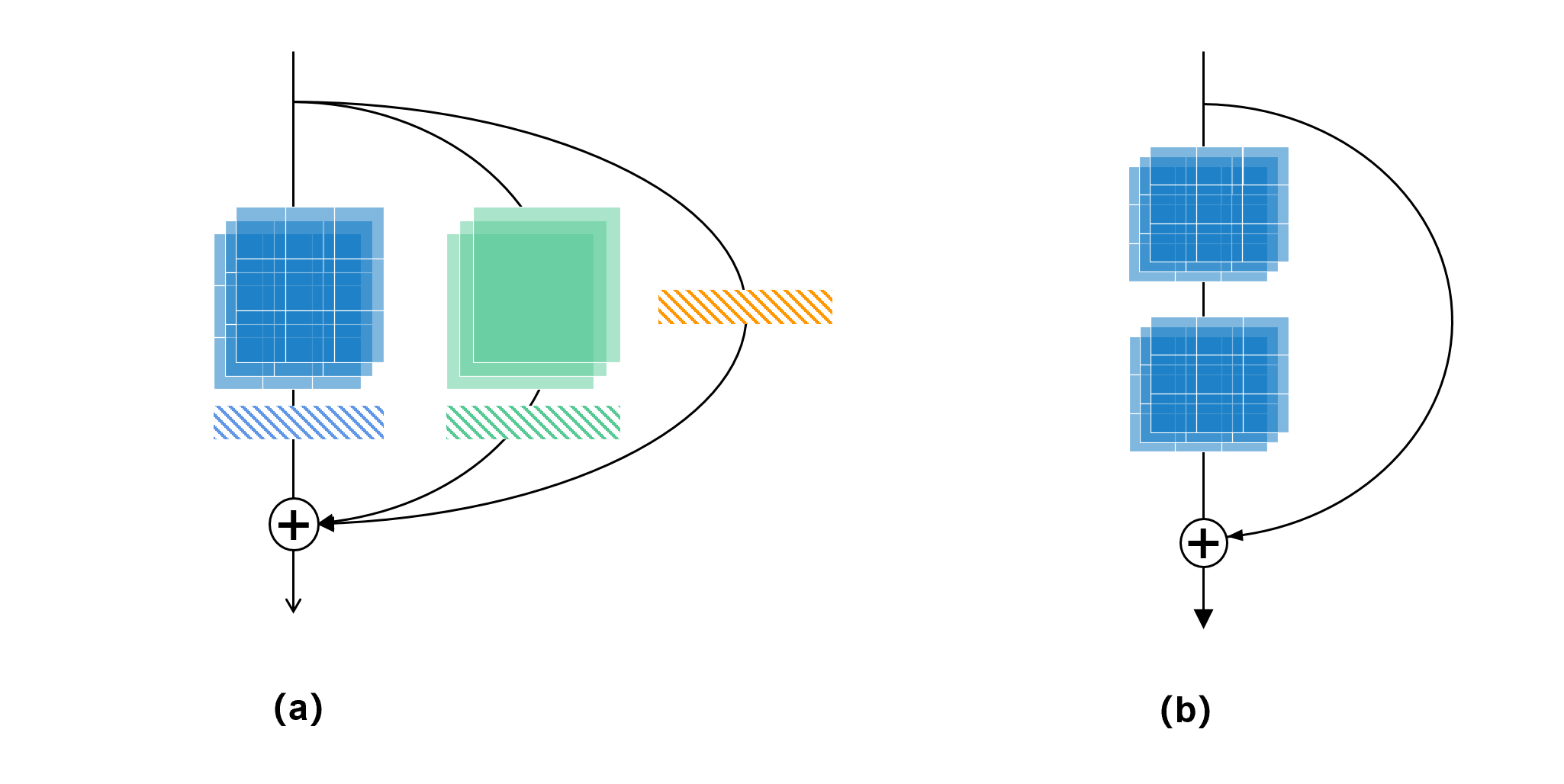}
  \caption{\textbf{Comparison between RepVGG and Basic Block.}}
  \label{rep_res}
\end{figure}

\subsubsection{Loss Function}
Recently, margin based softmax methods have been widely used in speaker recognition works. To make a much better performance, we strengthen the AM-Softmax \cite{wang2018additive, wang2018cosface} and AAM-Softmax \cite{deng2019arcface} loss functions by two methods.

First, the subcenter method \cite{deng2020sub} was introduced to reduce the influence of possible noisy samples. The formulation is given by:

\begin{equation}
  cos(\theta_{i,j})=\max_{1\leq k\leq K}(||x_i||\cdot||W_{j,k}||)
  \label{eq3}
\end{equation}
where the $\max$ function means that the nearest center is selected and it inhibits possible noisy samples interfering the dominant class center. 


Secondly, we proposed the Inter-TopK penalty to pay further attention to the centers which obtain high similarities comparing samples that do not belong to them. Therefore, it adds an extra penalty on these easily misclassified centers. Given a batch with $N$ examples and a number of classes of $C$, the formulation with adding extra Inter-TopK penalty based on the AM-Softmax is:


\begin{table*}[t]
  \centering
  \caption{\textbf{Ablation Study on Our Baseline System.} + here denotes stacking our methods.}
  \label{tab:tabel_one}
  \setlength{\tabcolsep}{1.5mm}{
  \begin{tabular}{lcccccccccc}
    \toprule
    \multirow{4}{*}{\textbf{Methods}} & 
    \multicolumn{2}{c}{\multirow{2}{*}{\textbf{VoxCeleb1-O}}} & 
    \multicolumn{2}{c}{\multirow{2}{*}{\textbf{VoxCeleb1-E}}} &  \multicolumn{2}{c}{\multirow{2}{*}{\textbf{VoxCeleb1-H}}} & 
    \multicolumn{2}{c}{\multirow{2}{*}{\textbf{VoxSRC20-dev}}} &  \multicolumn{2}{c}{\multirow{2}{*}{\textbf{VoxSRC21-val}}}   \\ \\
    \cline{2-11} 
    & \multirow{2}{*}[-2pt]{\textbf{EER(\%)}} &  \multirow{2}{*}[-2pt]{$\textbf{DCF}_\textbf{0.01}$} & \multirow{2}{*}[-2pt]{\textbf{EER(\%)}} & \multirow{2}{*}[-2pt]{$\textbf{DCF}_\textbf{0.01}$} & 
     \multirow{2}{*}[-2pt]{\textbf{EER(\%)}} & \multirow{2}{*}[-2pt]{$\textbf{DCF}_\textbf{0.01}$} & 
     \multirow{2}{*}[-2pt]{\textbf{EER(\%)}} & \multirow{2}{*}[-2pt]{$\textbf{DCF}_\textbf{0.05}$} &
     \multirow{2}{*}[-2pt]{\textbf{EER(\%)}} & \multirow{2}{*}[-2pt]{$\textbf{DCF}_\textbf{0.05}$} \\ \\
    \midrule
    ResNet-34-64      & 1.0660 & 0.0876   & 1.0440 & 0.0971   & 1.7660 & 0.1561   & 2.7580 & 0.1357   & 3.1300 & 0.1686 \\
    + K-Subcenter     & 0.9756 & 0.0840   & 1.0270 & 0.0930   & 1.7020 & 0.1467   & 2.6450 & 0.1321   & 2.7850 & 0.1503 \\
    ++ Inter-TopK    & 0.9730 & 0.0912   & 1.0170 & 0.0892   & 1.6860 & 0.1415   & 2.5760 & 0.1297   & 2.5800 & 0.1433 \\
    +++ MQMHA         & 0.9305 & 0.0738   & 0.9809 & 0.0879   & 1.6020 & 0.1373   & 2.5070 & 0.1246   & 2.5100 & 0.1403 \\
    ++++ Fine-tuning  & 0.6654 & 0.0573   & 0.8243 & 0.0725   & 1.3532 & 0.1154   & 2.1584 & 0.1054   & 1.9933 & 0.1158 \\
    +++++ AS-Norm     & 0.5594 & 0.0480   & 0.7632 & 0.0624   & 1.2122 & 0.0971   & 1.9120 & 0.0935   & 1.8367 & 0.0996 \\
    ++++++ QMF        & 0.5249 & 0.0498   & 0.7130 & 0.0627   & 1.1240 & 0.0923   & 1.8330 & 0.0867   & 1.6020 & 0.0906 \\
    \bottomrule
  \end{tabular}}
  \label{resnet_abl}
\end{table*}

\begin{equation}
\mathcal{L}_{AM'}\!=\!- \frac{1}{N}\!\sum_{i=1}^{N}log \frac{e^{s\cdot (cos\theta_{i,y_i}-m)}}
  {e^{s\cdot (cos\theta_{i,y_i}-m)}+\!\!\!\sum\limits_{j=1,j \neq y_i}^{C}e^{s\cdot \phi(\theta_{i,j})}}
  \label{eq4}
\end{equation}
where $m$ is the original margin of AM-Softmax and $s$ is the scalar of magnitude. We use the $\phi(\theta_{i,j})$ to replace the $cos\theta_{i,j}$ in the denominator:

\begin{equation}
  \phi(\theta_{i,j}) = 
  \begin{cases}
  cos\theta_{i,j}+m' &  j\in \mathop{arg\,topK}\limits_{1\leq n\leq C}(cos\theta_{i,n}) \\
  cos\theta_{i,j} & \textit{Others}.
  \end{cases}
  \label{eq5}
\end{equation}
where $m'$ is an extra penalty which focuses on the closest $K$ centers to the example $x_i$. And it is just the original AM-Softmax case when $m'=0$. Similarity, the Inter-TopK penalty could be also added for AAM-Softmax loss function by replacing $cos\theta_{i,j}+m'$ by $cos(\theta_{i,j}-m')$.

\subsection{Training Protocol}
We used Pytorch \cite{paszke2019pytorch} to conduct our experiments. All of our models were trained through two stages. 

In the first stage, the SGD optimizer with a momentum of 0.9 and weight decay of 1e-3 (4e-4 for online training) was used. We used 8 GPUs with 1,024 mini-batch and an initial learning rate of 0.08 to train all of our models. As described in section \ref{data_sec}, 200 frames of each sample in one batch were adopted to avoid over-fitting and speed up training. We adopted ReduceLROnPlateau scheduler with a frequency of validating every 2,000 iterations, and the patience is 2. The minimum learning rate is 1.0e-6, and the decay factor is 0.1. Furthermore, the margin gradually increases from 0 to 0.2 \cite{liu2019large}.

In the large-margin-based fine-tuning stage \cite{thienpondt2020idlab}, settings are slightly different from the first stage. Firstly, we removed the speed augmented part from the training set to avoid domain mismatch. Only 5,994 classes were left. Secondly, we changed the frame size from 200 to 600 and increased the margin exponentially from 0.2 to 0.5. The AM-Softmax loss was replaced by AAM-Softmax loss. The Inter-TopK penalty was removed to make training stable. Finally, We adopted a smaller finetuning learning rate of 8e-5 and a 256 batch size. The learning rate scheduler is almost the same while the decay factor became 0.5.

\subsection{Back-end}
After completing the fine-tuning stage, 512-dimensional speaker embeddings were extracted from the fully connected (FC) layer, and then the length normalization was applied before computing cosine similarity. Moreover, we utilized speaker-wise adaptive score normalization (AS-Norm) \cite{wang2020dku} and Quality Measure Functions (QMF) \cite{thienpondt2021idlab, thienpondt2020idlab} to calibrate the scores, and these methods greatly enhanced the performance. For AS-Norm, we selected the original VoxCeleb2 dev dataset without any augmentation. After extracting embeddings, all these embeddings were averaged speaker-wise, which resulted in 5994 cohorts. Then scores would be calibrated by this speaker-wise AS-Norm using top 400 imposter scores. For QMF, we combined three qualities, speech duration computed by Kaldi, imposter mean based on AS-Norm, and magnitude of non-normalized embeddings. Like IDLAB's way \cite{thienpondt2021idlab}, we also selected 30k trials from the original VoxCeleb2-dev as the training set of QMF. Then a Logistic Regression(LR) was trained to serve as our QMF model.

\begin{table}[h]
  \centering
  \caption{\textbf{Sub-System Structures.}}
  \label{tab:tabel_two}
  \setlength{\tabcolsep}{7.0mm}{
  \begin{tabular}{ll}
    \toprule
    \textbf{Index} & \textbf{Backbone} \\
    \midrule
    \emph{Offine fbank 81} \\
    \midrule
    \textbf{S1} & ResNet-34-64     \\
    \textbf{S2} & ResNet-101-64    \\
    \textbf{S3} & ResNet-152-64    \\
    \textbf{S4} & RepVGG-a2-96     \\
    \textbf{S5} & RepVGG-b1-64     \\
    \textbf{S6} & RepVGG-b2g4-64   \\
    \textbf{S7} & RepVGG-b2-64    \\
    \midrule
    \emph{Offine fbank 96} \\
    \midrule
    \textbf{S8} & RepVGG-b2g4-64  \\
    \midrule
    \emph{Online fbank 96} \\
    \midrule
    \textbf{S9} & RepVGG-b2g4-64  \\
    \bottomrule
  \end{tabular}}
  \label{subsys}
\end{table}

\begin{table*}[t]
  \centering
  \caption{\textbf{Results on Developing Sets.}}
  \label{tab:tabel_three}
  \setlength{\tabcolsep}{1.5mm}{
  \begin{tabular}{ccccccccccc}
    \toprule
    \multirow{4}{*}{\textbf{System Index}} & 
    \multicolumn{2}{c}{\multirow{2}{*}{\textbf{VoxCeleb1-O}}} & 
    \multicolumn{2}{c}{\multirow{2}{*}{\textbf{VoxCeleb1-E}}} &  \multicolumn{2}{c}{\multirow{2}{*}{\textbf{VoxCeleb1-H}}} & 
    \multicolumn{2}{c}{\multirow{2}{*}{\textbf{VoxSRC20-dev}}} &  \multicolumn{2}{c}{\multirow{2}{*}{\textbf{VoxSRC21-val}}}   \\ \\
    \cline{2-11} 
     & \multirow{2}{*}[-2pt]{\textbf{EER(\%)}} & \multirow{2}{*}[-2pt]{\textbf{$\textbf{DCF}_\textbf{0.01}$}} & \multirow{2}{*}[-2pt]{\textbf{EER(\%)}} & \multirow{2}{*}[-2pt]{\textbf{$\textbf{DCF}_\textbf{0.01}$}} & 
     \multirow{2}{*}[-2pt]{\textbf{EER(\%)}} & \multirow{2}{*}[-2pt]{\textbf{$\textbf{DCF}_\textbf{0.01}$}} & 
     \multirow{2}{*}[-2pt]{\textbf{EER(\%)}} & \multirow{2}{*}[-2pt]{\textbf{$\textbf{DCF}_\textbf{0.05}$}} &
     \multirow{2}{*}[-2pt]{\textbf{EER(\%)}} & \multirow{2}{*}[-2pt]{\textbf{$\textbf{DCF}_\textbf{0.05}$}} \\ \\
    \midrule
    \textbf{S1}   & 0.5249 & 0.0498   & 0.7130 & 0.0627   & 1.1240 & 0.0923   & 1.8330 & 0.0867   & 1.6020 & 0.0906 \\
    \textbf{S2}   & 0.5037 & 0.0356	 & 0.6435 & 0.0514   & \textbf{0.9737} & 0.0783   & 1.5760 & 0.0753   & \textbf{1.3350} & 0.0685 \\
    \textbf{S3}   & 0.4613 & \textbf{0.0232}	 & \textbf{0.6342} & \textbf{0.0477}   & 0.9932 & 0.0763   & \textbf{1.4770} & 0.0726   & 1.4550 & 0.0813 \\
    \textbf{S4}   & 0.5673 & 0.0309	 & 0.6759 & 0.0550	 & 1.0360 & 0.0830	 & 1.5860 & 0.0797	 & 1.4620 & 0.0776 \\
    \textbf{S5}   & \textbf{0.4401} & 0.0253	 & 0.6518 & 0.0494   & 0.9914 & 0.0738	 & 1.4960 & \textbf{0.0691}   & 1.3610 & \textbf{0.0628} \\
    \textbf{S6}   & 0.4825 & 0.0374   & 0.6707 & 0.0508   & 1.0270 & 0.0783   & 1.5160 & 0.0725   & 1.4050 & 0.0730 \\
    \textbf{S7}   & 0.4825 & 0.0283	 & 0.6511 & 0.0484	 & 0.9965 & 0.0738	 & 1.4910 & 0.0699	 & 1.4180 & 0.0660 \\
    \textbf{S8}   & 0.5090 & 0.0340	 & 0.6587 & 0.0489	 & 0.9954 & \textbf{0.0707}	 & 1.4940 & 0.0699	 & 1.4180 & 0.0698 \\
    \textbf{S9}   & 0.5673 & 0.0461	 & 0.6961 & 0.0584	 & 1.0910 & 0.0856	 & 1.7040 & 0.0845	 & 1.6420 & 0.0942 \\
    \midrule
    \emph{Fusion} \\
    \midrule
    \textbf{S1$\sim$S9}  & \textbf{0.4189}  &  \textbf{0.0217} & 	\textbf{0.5826}  &  \textbf{0.0414} & 	\textbf{0.8868} &   \textbf{0.0630}	 & \textbf{1.3400} &     \textbf{0.0624}	 & \textbf{1.2710} &    \textbf{0.0590} \\
    \bottomrule
  \end{tabular}}
  \label{sub_result}
\end{table*}

\subsection{Results}
\subsubsection{Baseline System Ablation Study}
In this subsection, we show our ablation study on our baseline system. The baseline system is a ResNet-34 backbone followed by MHA pooling and AM-Softmax. The performance was evaluated using the Equal Error Rate (EER) and the minimum Decision Cost Function (DCF) calculated where $C_{FA}=1$, $C_M=1$, and $P_{target}=0.01$ or $P_{target}=0.05$ for different trials. As Table \ref{resnet_abl} shows,  our baseline system's performance improved significantly on various trials by stacking our proposed methods gradually. For convenience, we took the performance of VoxSRC21-val as our benchmark. First, we conducted our ablation studies by changing normal AM-Softmax ($m=0.2, s=35$) to 3-subcenter AM-Softmax. The EER was improved from 3.13\% to 2.785\%, and the minDCF was improved from 0.1686 to 0.1503. By adding the Inter-TopK ($m'_{top5}=0.06$) extra penalty, the EER was 2.58\%, and the minDCF was 0.1433. Using MQMHA ($q=4, h=16$) instead of MHA, the EER further achieved 2.51\%,  and the minDCF was 0.1403. The procedures above already boosted our baseline system's EER by relatively 19.8\% and minDCF by relatively 16.78\%. The domain-based large margin finetuning improved our system performance from 2.51\% EER to 1.9933\% EER drastically. The minDCF also improved from 0.1403 to 0.1158. Applying the speaker-wise AS-Norm further achieved  1.8367\% EER and 0.0996 minDCF. The final QMF process got 1.60\% EER and 0.0906 minDCF. After doing AS-Norm and QMF, our system's EER improved 19.6\% relatively, and minDCF improved 21.76\% relatively compared to the finetuned system. After completing the ablation study, our baseline system improved EER relatively 48.9\%  and minDCF relatively 46.26\% in total.

For all our models, we followed the same procedure, and the only variable is our backbone.

\subsubsection{Sub-Systems and Fusion Performance}
All our sub-systems were described in Table \ref{subsys}. A total of 9 different backbones were used to generate different representations. The offline trained systems used both 81-dimensional and 96-dimensional acoustic features and online trained systems adopted 96-dimensional features only. Table \ref{sub_result} demonstrates the result achieved by our sub-systems of various trials. We found that a large model, such as RepVGG-B1, and ResNet-101 seemed to yield a better result compared to smaller models like our baseline system. However, an even bigger model like ResNet-152 and RepVGG-B2 could not bring a comparable performance boost regarding the drastically increased parameters. Also, it is worth mentioning that these even bigger models showed a sign of over-fitting on the VoxCeleb2-dev dataset. As the learning rate was smaller than 1e-4, the EER and minDCF of these large systems degraded. However, the performance of these systems remained SOTA even when we terminated the training at an earlier stage. 96-dimensional Fbank features were good complements of 81-dimensional Fbank features. The online system set we used is not the optimal choice, as we are still researching this new training paradigm. Though it shows a competitive result, it cannot achieve the best result of our large offline models.

Table \ref{eval_result} shows some of our submissions to the VoxSRC2021 and the final result of our fusion system. It is worth mentioning that our RepVGG-B1 achieved a 0.1212 minDCF and 2.2410\% EER with only a single model while ResNet-152 achieved a 0.1195 minDCF and 2.16\% EER. We tuned our fusion weights of all these models based on the results of VoxCeleb1-H and VoxSRC21-val. The final fusion resulted in a 0.1034 minDCF and a 1.846\% EER in the VoxSRC2021 challenge. The fusion result improved 12.47\% relatively in minDCF and 14.54\% relatively in EER compared to our ResNet-152 model.

\begin{table}[h]
  \centering
  \caption{\textbf{Our Submissions to VoxSRC21-test.}}
  \label{tab:tabel_four}
  \begin{tabular}{ccc}
    \toprule
    \textbf{System Index} & \textbf{EER(\%)} & $\textbf{DCF}_{\textbf{0.05}}$  \\ 
    \midrule
    \textbf{S1}   & 2.8890 & 0.1700  \\
    \textbf{S2}   & - & -  \\
    \textbf{S3}   & 2.1690 & 0.1195 \\
    \textbf{S4}   & - & -  \\
    \textbf{S5}   & 2.2410 & 0.1212  \\
    \textbf{S6}   & - & -  \\
    \textbf{S7}   & - & -  \\
    \textbf{S8}   & - & -  \\
    \textbf{S9}   & - & -  \\
    \midrule
    \emph{Fusion} \\
    \midrule
    \textbf{S1$\sim$S9}  & \textbf{1.8460} &  \textbf{0.1034} \\
    \bottomrule
  \end{tabular}
  \label{eval_result}
\end{table}

\section{Conclusions}
In this challenge, we first introduced a new backbone structure (RepVGG) in speaker verification. We also proposed MQMHA, Inter-TopK loss, and domain-based large margin fine-tuning methods. All these methods above and the large backbones ensured our first place in track 1 and track 2 of VoxSRC 2021. The final result of our system was 0.1034 minDCF and 1.846\% EER.

\section{Acknowledgements}
This work is supported by the SpeakIn Technologies Co. Ltd.


\bibliographystyle{IEEEtran}
\end{document}